 \documentclass[twocolumn]{aastex6}

\begin{document}


\title{The First Spectrum of the Coldest Brown Dwarf}

\author{Andrew J. Skemer\altaffilmark{1}, 
Caroline V. Morley\altaffilmark{1}, 
Katelyn N. Allers\altaffilmark{2}, 
Thomas R. Geballe\altaffilmark{3}, 
Mark S. Marley\altaffilmark{4}, 
Jonathan J. Fortney\altaffilmark{1}, 
Jacqueline K. Faherty\altaffilmark{5,6,7}, 
Gordon L. Bjoraker\altaffilmark{8}, 
Roxana Lupu\altaffilmark{4}}

\altaffiltext{1}{University of California, Santa Cruz}
\altaffiltext{2}{Bucknell University}
\altaffiltext{3}{Gemini Observatory}
\altaffiltext{4}{NASA Ames Research Center}
\altaffiltext{5}{Carnegie, Department of Terrestrial Magnetism}
\altaffiltext{6}{American Museum of Natural History}
\altaffiltext{7}{Hubble Fellow}
\altaffiltext{8}{NASA Goddard Space Flight Center}

\begin{abstract}

The recently discovered brown dwarf WISE 0855 presents our first opportunity to directly study an object outside the Solar System that is nearly as cold as our own gas giant planets. However the traditional methodology for characterizing brown dwarfs---near infrared spectroscopy---is not currently feasible as WISE 0855 is too cold and faint. To characterize this frozen extrasolar world we obtained a 4.5-5.2 $\mu$m spectrum, the same bandpass long used to study Jupiter's deep thermal emission. Our spectrum reveals the presence of atmospheric water vapor and clouds, with an absorption profile that is strikingly similar to Jupiter.  The spectrum is high enough quality to allow the investigation of dynamical and chemical processes that have long been studied in Jupiter's atmosphere, but now on an extrasolar world.

\end{abstract}

\keywords{stars: brown dwarfs --- 
stars: atmospheres  --- planets and satellites: atmospheres --- planets and satellites: individual: WISE 0855 }



\section{Introduction} \label{sec:intro}
The coldest characterizable exoplanets are still much hotter than the gas giant planets in our Solar System \citep{2013ApJ...774...11K,2015Sci...350...64M}.  However, with the recent discovery of a $\sim$250 K brown dwarf, WISE 0855, we now have our first opportunity to directly study an object whose physical characteristics are similar to Jupiter \citep{2014ApJ...786L..18L}.  WISE 0855 is the nearest known planetary mass object, and the coldest known compact object outside of our Solar System \citep{2014ApJ...786L..18L}.  Its extremely low temperature makes it the first object after Earth, Mars, Jupiter and Saturn likely to host water clouds in its visible atmosphere \citep{1999ApJ...513..879M,2000ApJ...538..885S,2014ApJ...793L..16F}.

A handful of photometric detections and non-detections confirm that W0855 is cold but only allow us to speculate on its atmospheric composition \citep{2014ApJ...793L..16F,2014AJ....148...82W,2014ApJ...797....3K,2014A&A...570L...8B}.  WISE 0855 is too faint to characterize with conventional spectroscopy in the optical or near infrared ($<$2.5 $\mu$m).  Like Jupiter, a minimum in $\rm CH_{4}$ and $\rm H_{2}$ gas opacity allows thermal emission from the deep atmosphere to escape through a 5 $\mu$m atmospheric window \citep{1969BAAS....1..200L,1986Icar...66..579B,1986ApJ...311.1058B}.  Spectroscopy at these wavelengths is challenging but not impossible.  While there are currently no space-based facilities capable of 5 $\mu$m spectroscopy, ground-based telescopes can observe through the Earth's 4.5-5.2 $\mu$m atmospheric window, although sensitivity is limited by the brightness of the Earth's sky.  With careful calibration, we were able to obtain a spectrum of WISE 0855, an object that is 5 times fainter than the faintest object previously detected with ground-based 5 $\mu$m spectroscopy \citep{2014AJ....148...82W,2009ApJ...695..844G}.

\section{Observations}
We observed WISE 0855 with the Gemini-North telescope and the Gemini Near Infrared Spectrograph \citep[GNIRS; ][]{2006SPIE.6269E..4CE}.  Gemini-North is located near the summit of Mauna Kea, a cold location that also provides some of the driest conditions of any astronomical site in the world.  Because of those attributes as well as the telescope's low emissivity, the infrared background incident on Gemini's instruments is the lowest of any 8-10 meter class telescope. Additionally, Gemini-North is operated in queue-mode, allowing us to observe WISE 0855 over many nights in clear, dry and calm conditions.  In total, we observed WISE 0855 for 14.4 hours over 13 nights.  We calibrated the transmission of the Earth's atmosphere by observing standard stars before and after every observation of WISE 0855.

GNIRS was configured with its long-wavelength camera, 31.7 line/mm grating and 0.675'' slit, which provides a single order spectrum covering the M-band (4.5-5.2 $\mu$m) at a spectral resolution of R$\sim$800.  Although GNIRS is typically used with its fastest readout mode for broad M-band spectroscopy, we found that there was still appreciable read-noise at some wavelengths, and decided to operate in its 2nd fastest mode, which uses 16 digital averages and has an overhead of 0.6 seconds per frame.  In order to keep our observations of WISE 0855 efficient, we integrated for 2.5 seconds, which is enough to saturate the brightest M-band sky-lines.  These wavelengths were then masked in our later analysis.

Observing conditions in the Gemini queue were limited to photometric conditions, $<$0.8'' seeing at V-band, $<$3 mm precipitable water vapor, and $<$1.5 airmass.  Telluric calibrators (HIP 39898 and HIP 49900) were obtained at similar airmasses before and after each observation, with a maximum separation of 2 hours.

Because WISE 0855 is essentially invisible in the near-infrared \citep{2014ApJ...793L..16F,2014ApJ...797....3K,2014A&A...570L...8B}, acquisitions were done using a blind offset from a nearby star with a recent astrometric calibration \citep{2014ApJ...793L..16F}.  The position of WISE 0855 was calculated by propagating its proper motion and parallax \citep{2014AJ....148...82W,2014ApJ...796....6L}.  Gemini's offset pointing is accurate to $\sim$0.2'', and our position estimate for WISE 0855 is accurate to $\sim$0.1''.  Thus, it is possible that WISE 0855 was slightly misaligned with the slit.  However, we chose an oversized slit (0.675'') to mitigate the effects of a slight misalignment.  To enable subtraction of the thermal background, we took observations in an ABBA pattern, nodding 6'' along the slit. A dataset of WISE 0855 consists of 9 ABBA sequences for a total of 36 images and 36 minutes of integration time.  In total, we obtained 24 of these datasets for a combined integration time of 14.4 hours.  A summary of the dates, conditions, and telluric calibrators for our observations is listed in Table 1.

\begin{deluxetable}{lcccccc}
\tabletypesize{\scriptsize}
\tablecaption{Observations}
\tablewidth{0pt}
\tablehead{
\colhead{Date} &
\colhead{WISE 0855} &
\colhead{IQ} &
\colhead{WV} &
\colhead{Telluric} &
\colhead{Telluric}
\\
\colhead{} &
\colhead{Airmass} &
\colhead{} &
\colhead{} &
\colhead{} &
\colhead{Airmass}
}
\startdata
2015-12-01 & 1.29 & 20\% & 50\% & HIP 39898 & 1.35 \\
2015-12-01 & 1.16 & 20\% & 50\% & HIP 49900 & 1.22 \\
2015-12-01 & 1.14 & 20\% & 50\% & HIP 49900 & 1.13 \\
2015-12-03 & 1.26 & 20\% & 50\% & HIP 39898 & 1.31 \\
2015-12-03 & 1.15 & 20\% & 50\% & HIP 49900 & 1.20 \\
2015-12-05 & 1.13 & 20\% & 20\% & HIP 39898 & 1.17 \\
2015-12-05 & 1.15 & 20\% & 20\% & HIP 49900 & 1.12 \\
2015-12-06 & 1.18 & 70\% & 20\% & HIP 39898 & 1.22 \\
2015-12-06 & 1.13 & 70\% & 20\% & HIP 49900 & 1.14 \\
2015-12-07 & 1.13 & 20\% & 20\% & HIP 39898 & 1.17 \\
2015-12-14 & 1.32 & 70\% & 20\% & HIP 39898 & 1.39 \\
2015-12-14 & 1.17 & 70\% & 20\% & HIP 49900 & 1.24 \\
2015-12-29 & 1.14 & 70\% & 20\% & HIP 39898 & 1.19 \\
2015-12-29 & 1.13 & 70\% & 20\% & HIP 49900 & 1.13 \\
2015-12-30 & 1.14 & 70\% & 20\% & HIP 39898 & 1.19 \\
2015-12-30 & 1.13 & 70\% & 20\% & HIP 49900 & 1.13 \\
2015-12-30 & 1.28 & 70\% & 20\% & HIP 49900 & 1.13 \\
2015-12-31 & 1.23 & 70\% & 20\% & HIP 39898 & 1.30 \\
2016-01-01 & 1.29 & 20\% & 20\% & HIP 39898 & 1.38 \\
2016-01-02 & 1.25 & 70\% & 20\% & HIP 39898 & 1.29 \\
2016-01-03 & 1.29 & 70\% & 20\% & HIP 39898 & 1.33 \\
2016-01-03 & 1.19 & 70\% & 20\% & HIP 49900 & 1.32 \\
2016-01-05 & 1.26 & 70\% & 20\% & HIP 39898 & 1.34 \\
2016-01-05 & 1.15 & 70\% & 20\% & HIP 49900 & 1.20
\enddata
\tablecomments{IQ and WV refer to Gemini's image quality and precipitable water vapor observational constraint boundaries (\url{https://www.gemini.edu/node/10781})}
\label{observations}
\end{deluxetable}

\section{Reductions}
We reduced the data using a modified version of the REDSPEC package\footnote{\url{http://www2.keck.hawaii.edu/inst/nirspec/redspec}} to spatially and spectrally rectify each exposure with the dispersion direction as ``rows'' and the spatial direction as ``columns''. 

For each dataset, we first create a non-linearity flag image by marking pixels with values greater than 10,000 counts per coadd.  We then median scale the images and create nod-subtracted A-B pairs.

Observations of the telluric standards are used to create the spatial and spectral maps, which are then used to rectify the WISE 0855 data.  The spatial rectification process uses a subtracted A-B pair of telluric star exposures, fits gaussians to find the center of each star as a function of wavelength, and then fits a 2nd order polynomial to define a trace.  These traces are then used to re-map each exposure so that the dispersion direction lies along detector rows.  To wavelength calibrate and spectrally rectify our exposures, we use sky emission lines in our telluric standard spectra.  We determined the wavelengths of 11 sky lines by smoothing a model of the sky background for Mauna Kea\footnote{\url{http://www.gemini.edu/sciops/telescopes-and-sites/}\\\url{observing-condition-constraints/ir-background-spectra}} to match the R$\sim$800 resolution of our GNIRS spectra.  We use a 2nd order polynomial to fit a dispersion solution to each spectral trace, which is then used to remap each exposure so that each column corresponds to a single wavelength.  The residuals for our dispersion fits are $<$2 pixels or 0.0013 $\mu$m.

For each rectified A-B pair image, we remove residual skylines from each column by subtracting that column's median value.  We combine the rectified, background-subtracted A-B pairs with a 3-$\sigma$ clipped mean at each pixel. The uncertainty of each pixel in the stacked image is the standard deviation of the mean. 
 
To extract the spectrum, we create a wavelength-collapsed spatial cut by taking the median of each row in the stacked A-B image, and boxcar smoothing the profile by 5 pixels (0.25'').  We identify the location of the A and B spectra by fitting two Gaussians to the spatial cut.  We then extract each spectrum using an aperture radius of $\sim$1.6$\sigma$ (where $\sigma$ is the width of the Gaussian fit to the spatial cut), and subtract any residual sky background by fitting a line to the background regions.  If a pixel that was flagged as non-linear falls in the aperture, the spectrum at that wavelength is also flagged.  For each dataset, our reduction results in an A and a B spectrum.  We extract the spectra of our telluric standard stars using a similar process.

We divide each extracted spectrum of WISE 0855 by the spectrum of an adjacent telluric standard, and multiply by an appropriate blackbody.  When using the calibrator HIP 39898, there is an additional step to remove its Pfund $\beta$ emission by fitting and subtracting a Gaussian. In most cases, the airmass of our WISE 0855 observation matched the airmass of the telluric observation to within 0.1.  In total, we have 48 telluric calibrated spectra of WISE 0855, which we combine using a robust weighted mean \citep{2004PASP..116..362C}.  

\begin{figure}
\begin{center}
\hspace{-10 mm}
\includegraphics[angle=0,width=1.1\columnwidth]{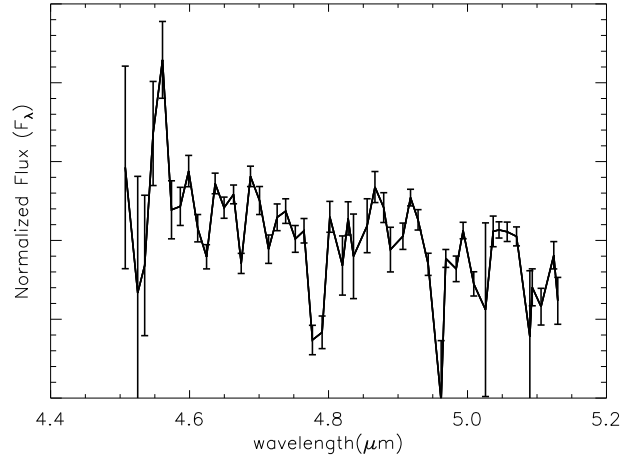}
\caption{Gemini/GNIRS spectrum of WISE 0855 showing numerous absorption features.  The spectrum is normalized due to the flux calibration uncertainty associated with blind-offset slit misalignment.  
\label{spectrum}}
\end{center}
\end{figure}

We binned the data in 20 pixel increments for presentation in this paper (spectral resolution of $\sim$300 per point, or $\sim$150 for Nyquist sampling).  Within each bin, we calculated the weighted average and weighted error, which has the effect of suppressing wavelengths with large telluric emission (and low transmission) features.  Our calibrated spectrum of WISE 0855 is presented in Figure \ref{spectrum}.  For subsequent spectrum plots in this paper, identical weights are applied to comparison spectra, and are used to calculate each bin's effective wavelength.

In addition to making the spectrum intelligible, binning the data smooths wavelength mismatch errors between WISE 0855 and the telluric calibrators, which would result from WISE 0855 being slightly misaligned in the oversized slit.  To confirm that the wavelength mismatch errors did not affect our final spectrum, we re-reduced the data assuming a 0.2'' slit misalignment for WISE 0855 and found negligible differences.

\begin{figure}
\begin{center}
\hspace{-10 mm}
\includegraphics[angle=90,width=1.1\columnwidth]{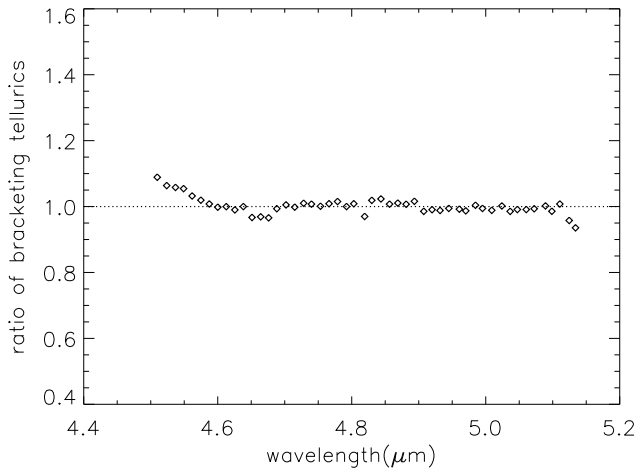}
\caption{Ratio of the telluric standard spectra observed before and after WISE 0855.  From 4.6-5.1 $\mu$m the spectrum is flat with a standard deviation of 2.5\%, implying that the errors in the telluric correction are negligible compared to the photon-noise dominated errors in our WISE 0855 spectrum.  From 4.5-4.6 $\mu$m and 5.1-5.15 $\mu$m, the telluric calibration is slightly worse, but still smaller than the reported errors in the WISE 0855 spectrum.  The data in this figure are binned with the same weighted average as the WISE 0855 spectra, and have been normalized to 1. 
\label{telluric}}
\end{center}
\end{figure}

The overall quality of our data reduction process and telluric correction is validated by dividing the spectra of our pre and post observation telluric calibrators, averaged over the 13 nights of observations.  Figure \ref{telluric} shows that the ratio of the telluric standards is flat to within 2.5\% from 4.6-5.1 $\mu$m, with slightly larger errors from 4.5-4.6 $\mu$m and 5.1-5.15 $\mu$m.  In all cases, the telluric mismatch is negligible compared to the background photon noise seen in our spectrum.

\section{Description of Atmosphere Models}
We interpret the observed spectrum of WISE 0855 using custom atmospheric models.  All of our models have an effective temperature of 250 K, which is well-constrained from WISE 0855's measured luminosity and a radius estimate that is relatively insensitive to unknown parameters, such as WISE 0855's age \citep{2014ApJ...786L..18L}.

We calculate cloud-free 1D pressure-temperature profiles assuming both chemical and radiative-convective equilibrium. The thermal radiative transfer is determined using the ``source function technique'' presented in \citet{1989JGR....9416287T}. The gas opacity is calculated using correlated-k coefficients; our opacity database is described extensively in \citet{2008ApJS..174..504F} and \citet{2014ApJS..214...25F}. The atmosphere models are more extensively described in \citet{1989Icar...80...23M,1996Sci...272.1919M,1999ApJ...513..879M,2002ApJ...568..335M,2008ApJ...689.1327S,2008ApJ...678.1419F}.

Exoplanets and brown dwarfs cooler than $\sim$350 K are expected to form water ice clouds in their upper atmospheres which are optically thick enough to alter the emergent spectrum \citep{1999ApJ...513..879M,2000ApJ...538..885S,2003ApJ...596..587B,2014ApJ...787...78M}.  We find that partly cloudy models \citep{2014ApJ...787...78M} are indistinguishable from cloud-free models after normalizing and binning over the wavelength range of our WISE 0855 spectrum.   Radiative-convective equilibrium models with full water cloud coverage do not converge well between $T_{eff}\sim$200-450 K \citep{2014ApJ...787...78M}. 

For this reason, we use a simplified approach to consider the effect of clouds at different layers, following studies of Jupiter's mid-infrared spectral features \citep{2015ApJ...810..122B}. We calculate the pressure-temperature profile of a cloud-free atmosphere in radiative-convective equilibrium. We then use that profile and calculate the thermal emission using an adapted version of {\tt DISORT} \citep{2015ApJ...815..110M} including a gray, fully absorbing cloud with optical depth $\tau = $1. We vary the cloud-top pressure of this gray cloud to best match the observed spectrum of WISE 0855.
 
We also use the {\tt DISORT} radiative transfer scheme to model the effect of non-equilibrium abundances of trace gases such as $\rm PH_{3}$ and $\rm CH_{3}D$.  We first calculate a cloud-free profile in chemical equilibrium, and then calculate spectra in which we change the abundance of each gas separately, changing the mixing ratio of that gas to be uniform at that value throughout the atmosphere.

\begin{figure}
\begin{center}
\vspace{-25 mm}
\includegraphics[angle=0,width=1.1\columnwidth]{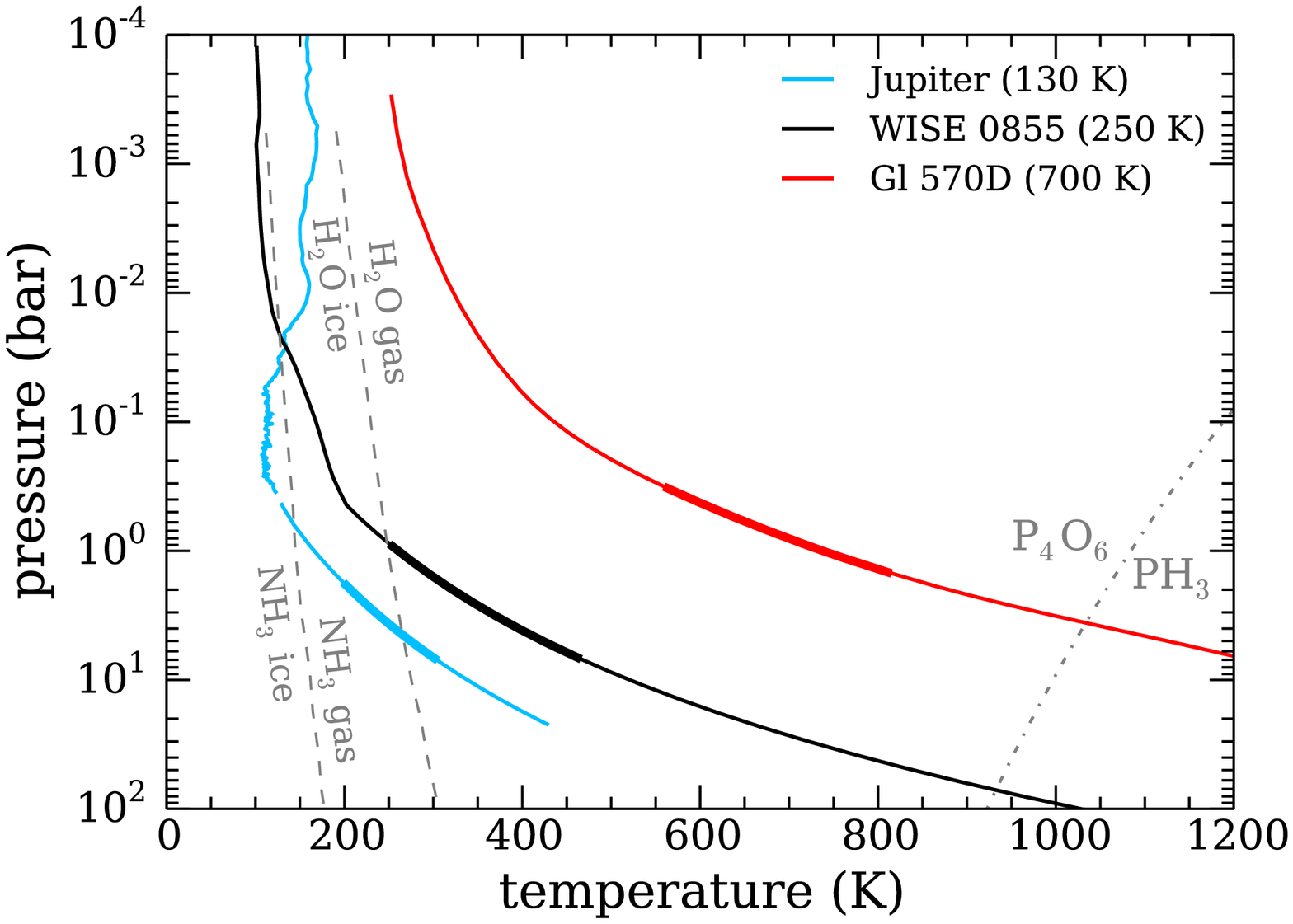}
\vspace{-30 mm}
\caption{Temperature-pressure profiles of Jupiter \citep{1998JGR...10322857S}, WISE 0855, and what was previously the coldest extrasolar object with a 5 $\mu$m spectrum, Gl 570D \cite{2012ApJ...760..151S}.  The approximate locations of each object's 5 $\mu$m photosphere are denoted with thicker lines (for Jupiter, the photosphere is as seen through cloud-free hot-spots).  Two dashed lines show the boundaries where $\rm H_{2}O$ gas and $\rm NH_{3}$ gas begin to condense into clouds composed of $\rm H_{2}O$ ice and $\rm NH_{3}$ ice.  A dot-dash line divides the regions where phosphorus is primarily in $\rm P_{4}O_{6}$ versus $\rm PH_{3}$ under equilibrium chemistry assumptions. WISE 0855 and Jupiter have relatively similar temperature-pressure profiles, with photospheres that are in the vicinity of the condensation points for $\rm NH_{3}$ and $\rm H_{2}O$.
\label{P-T}}
\end{center}
\end{figure}

Our model temperature-pressure profile for WISE 0855 and the next coldest brown dwarf with a 5 $\mu$m spectrum \citep[Gl 570D; 700K][]{2012ApJ...760..151S} are shown in Figure \ref{P-T}, along with a measured profile for Jupiter \citep{1998JGR...10322857S} extrapolated along an adiabat to lower depths.  Jupiter and WISE 0855 have relatively similar temperature-pressure profiles, thus their photospheres should display similar features.  Both photospheres are near the condensation temperatures for $\rm H_{2}O$ and $\rm NH_{3}$ clouds.  Under equilibrium chemistry assumptions, neither photosphere should have detectable amounts of $\rm PH_{3}$.

\section{Discussion}  
\subsection{Composition and Clouds}
\begin{figure*}
\begin{center}
\includegraphics[angle=0,width=2\columnwidth]{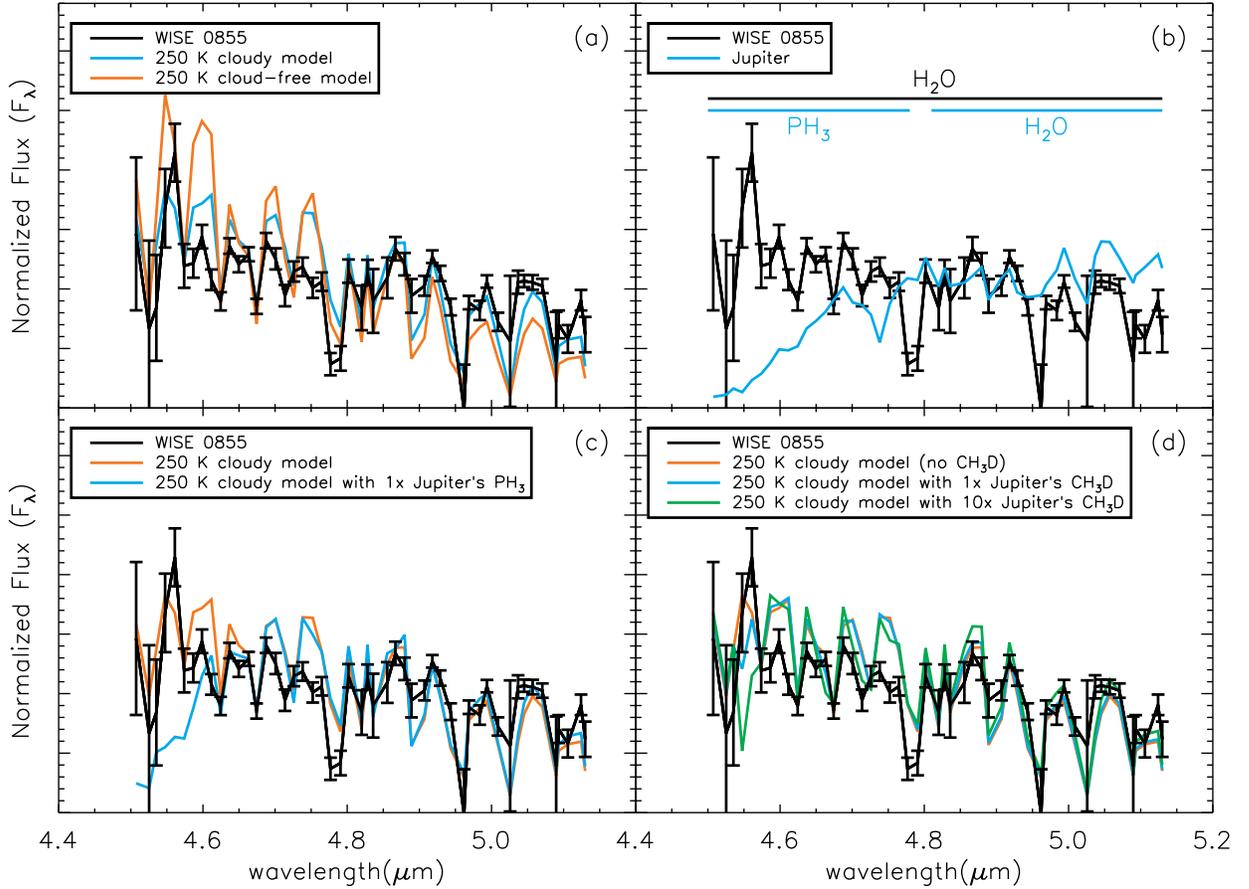}
\vspace{10 mm}
\caption{(a) Gemini/GNIRS spectrum of WISE 0855 compared to 250 K cloudy and cloud-free brown dwarf models.  The locations of the absorption features match well between the data and  models, and are all consistent with water vapor.  The shapes of the absorption features are affected by clouds (likely $\rm H_{2}O$), which mute the spectrum providing a better fit between the model and the data than the cloud-free case.  The remaining discrepancies between the cloudy model and the data are likely due, in part, to additional complexities (such as 3-dimensional structures) of the clouds. (b) Gemini/GNIRS spectrum of WISE 0855 compared to a full-disk spectrum of Jupiter from the Infrared Space Observatory \citep[ISO;][]{1996A&A...315L.397E}, binned to WISE 0855's resolution.  The spectra show striking similarities from 4.8-5.2 $\mu$m (with a notable exception at 4.96 $\mu$m), as both objects are dominated by $\rm H_{2}O$ \citep{1982ApJ...263..443K}.  From 4.5-4.8 $\mu$m, Jupiter's spectrum is dominated by $\rm PH_{3}$, which has been used to infer turbulent diffusion of hot material from deep in Jupiter's atmosphere \citep{1982ApJ...263..443K,1975Sci...190..274P,2006ApJ...648.1181V}.  The WISE 0855 spectrum does not show a similar $\rm PH_{3}$ feature. (c) Gemini/GNIRS spectrum of WISE 0855 compared to models with equilibrium chemistry $\rm PH_{3}$, and enhanced $\rm PH_{3}$ to match the abundance measured in Jupiter.  The WISE 0855 spectrum is strongly inconsistent with the enhanced $\rm PH_{3}$ model, suggesting that WISE 0855 does not have the same turbulent mixing seen on Jupiter. (d) Gemini/GNIRS spectrum of WISE 0855 compared to models that vary $\rm CH_{3}D$ abundance.  The WISE 0855 spectrum has the raw sensitivity to measure $\rm CH_{3}D$ at 4.55 $\mu$m, but the imperfect match between the rest of the spectrum and our current model would hamper an analysis.
\label{four panel}}
\end{center}
\end{figure*}

Our equilibrium chemistry models from Section 4 predict that for a 250 K object, all of the major absorption features from 4.5-5.2 $\mu$m are the result of water vapor.  Figure \ref{four panel}a shows cloudy and cloud-free models compared to the WISE 0855 spectrum.  The wavelengths of the model features match the wavelengths of the spectrum features, which suggests that our WISE 0855 spectrum is dominated by water vapor. 

The cloudy model is a better fit to the depths of WISE 0855's absorption features, as well as its overall slope.  While our model does not specify a particular cloud composition, WISE 0855 is at a temperature where the clouds are likely to be composed of water or water ice \citep{2003ApJ...596..587B,2012ApJ...756..172M}.  The cloudy model's fit to WISE 0855 is imperfect, which suggests that there is additional complexity that is beyond the scope of our current models.  For example, the particular vertical and horizontal structure of the clouds is likely to impact WISE 0855's appearance.

\subsection{Comparison with Jupiter}
We compare our spectrum of WISE 0855 to a full-disk spectrum of Jupiter in Figure \ref{four panel}b.  From 4.8-5.2 $\mu$m, WISE 0855 and Jupiter are strikingly similar, as Jupiter's atmosphere is also dominated by $\rm H_{2}O$ absorption at these wavelengths \citep{1982ApJ...263..443K}.  From 4.5-4.8 $\mu$m, Jupiter's spectrum is dominated by $\rm PH_{3}$ (phosphine) absorption \citep{1982ApJ...263..443K}.  If Jupiter were in chemical equilibrium, phosphorus would exist in the form of $P_{4}O_{6}$ in its photosphere and $\rm PH_{3}$ in its hotter interior \citep{1975Sci...190..274P,2006ApJ...648.1181V}.  The existence of $\rm PH_{3}$ in Jupiter's spectrum is evidence that Jupiter's atmosphere turbulently mixes gases from its hot interior into its cooler photosphere on a faster timescale than the $\rm PH_{3} \rightarrow P_{4}O_{6}$ reaction sequence can reach equilibrium \citep{1975Sci...190..274P,2006ApJ...648.1181V}.
 
\subsection{$\rm PH_{3}$ Chemistry and Mixing}
WISE 0855's spectrum does not show the strong $\rm PH_{3}$ absorption seen in Jupiter's spectrum.  In Figure \ref{four panel}c, we show our equilibrium chemistry model for WISE 0855, along with a model that has $\rm PH_{3}$ enhanced to the abundance measured in Jupiter \citep{Lodders2010}.  If WISE 0855 had Jupiter's abundance of $\rm PH_{3}$, it would easily be visible in our spectrum.  The fact that WISE 0855 has less $\rm PH_{3}$ than Jupiter most likely implies that WISE 0855 has less turbulent mixing than Jupiter, although atmospheric metallicity and gravity may also play a role. 

\subsection{Deuterium}
Figure \ref{four panel}d shows a comparison of WISE 0855 with models that vary $\rm CH_{3}D$ (deuterated methane).  Deuterium is expected to be quickly depleted by fusion in objects more massive than $\sim$13 M$\rm _{jup}$, a boundary commonly used to separate planets from brown dwarfs \citep{1997ApJ...491..856B,2011ApJ...727...57S}.  Thus $\rm CH_{3}D$ measurements can be used to glean information about the masses of exoplanets and free-floating planets/brown-dwarfs.  In our Solar System, deuterium becomes chemically concentrated in certain environments, and has been used to study the formation of the giant planets, and the delivery of water to Earth \citep{1986Natur.320..244O,2015Sci...347A.387A}. In multi-planet extrasolar systems, $\rm CH_{3}D$ measurements could similarly be used to study the composition of the planets' nascent materials. 

Our spectrum has the sensitivity to distinguish between models with Jupiter's approximately primordial deuterium abundance \citep{1986Natur.320..244O,Lodders2010}, and models with no deuterium.  However, the $\rm CH_{3}D$ absorption feature is blended with water absorption features, which need to be better understood before we can really measure $\rm CH_{3}D$. 

\section{Future Prospects}
To understand the nature of clouds and vertical mixing in cold gas giants, we need 5 $\mu$m spectra of objects across a continuum of temperatures.  After Jupiter (130 K) and WISE 0855 (250 K), the next coldest object with a 5 $\mu$m spectrum is Gl 570D (700 K) \citep{2012ApJ...760..151S}.  In the near term, ground-based telescopes have the sensitivity to obtain spectra of a handful of objects in this temperature range.  Future facilities, like the James Webb Space Telescope (JWST), will have the sensitivity to characterize cold gas giants with higher precision, and at wavelengths not possible from Earth \citep{2014ApJ...787...78M}.  The next generation of ground-based telescopes (Extremely Large Telescopes, or ELTs) will have the angular resolution and sensitivity to study systems with multiple exoplanets that look like WISE 0855.  With spectrographs operating in the 5 $\mu$m atmospheric window \citep{2015SPIE.9605E..1DS}, it will be possible to compare large samples of exoplanets at wavelengths that have revealed much of what we know about gas giants in our own Solar System.

\acknowledgements
Based on observations obtained at the Gemini Observatory, which is operated by the Association of Universities for Research in Astronomy, Inc., under a cooperative agreement with the NSF on behalf of the Gemini partnership: the National Science Foundation (United States), the National Research Council (Canada), CONICYT (Chile), Ministerio de Ciencia, Tecnolog\'{i}a e Innovaci\'{o}n Productiva (Argentina), and Minist\'{e}rio da Ci\^{e}ncia, Tecnologia e Inova\c{c}\~{a}o (Brazil).  The authors thank Mike Cushing and Pat Irwin for supplying compiled spectroscopy of Jupiter, and Satoko Sorahana for supplying AKARI spectroscopy of T dwarfs.
\\\\
\bibliographystyle{aasjournal}

\end{document}